\documentclass[conference,a4paper]{IEEEtran}
\usepackage{graphicx,xcolor,tikz}
\usepackage{subfig}
\usepackage{float}
\usepackage{cite}
\usepackage{makecell}

\newcommand{\ignore}[1]{}

\usepackage[left=1.41cm,right=1.44cm,top=1.91cm,bottom=4.2cm,bindingoffset=0cm]{geometry}
\usepackage{multirow}
\DeclareRobustCommand*{\IEEEauthorrefmark}[1]{\raisebox{0pt}[0pt][0pt]{\textsuperscript{\footnotesize #1}}}

\begin{document}

\title{Lightweight UAV-based Measurement System 
for Air-to-Ground Channels at 28 GHz}

\author{\IEEEauthorblockN{
Vasilii Semkin\IEEEauthorrefmark{1},   
Seongjoon Kang\IEEEauthorrefmark{2},   
Jaakko Haarla\IEEEauthorrefmark{3}, 
William Xia\IEEEauthorrefmark{2},    
Ismo Huhtinen\IEEEauthorrefmark{1},      
Giovanni Geraci\IEEEauthorrefmark{4},\\
Angel Lozano\IEEEauthorrefmark{4},
Giuseppe Loianno\IEEEauthorrefmark{2},
Marco Mezzavilla\IEEEauthorrefmark{2},
and Sundeep Rangan\IEEEauthorrefmark{2}
}                                     
\IEEEauthorblockA{\IEEEauthorrefmark{1}
VTT Technical Research Centre of Finland, (\textit{vasilii.semkin@vtt.fi})}
\IEEEauthorblockA{\IEEEauthorrefmark{2}
NYU Tandon School of Engineering, Brooklyn, NY, USA}
\IEEEauthorblockA{\IEEEauthorrefmark{3}
Aalto University, School of Electrical Engineering,  Finland}
\IEEEauthorblockA{\IEEEauthorrefmark{4}
Univ. Pompeu Fabra, Barcelona, Spain}
}

\maketitle

\begin{abstract}
Wireless communication at millimeter wave frequencies has attracted considerable attention for the delivery of high-bit-rate connectivity to unmanned aerial vehicles (UAVs). However, conducting the channel measurements necessary to assess communication at these frequencies has been challenging due to the severe payload and power restrictions in commercial UAVs.
This work presents a novel lightweight (approximately {1.3}~{kg}) channel measurement system at {28}~{GHz} installed on a commercially available UAV. A ground transmitter equipped with a horn antenna
conveys sounding signals to a UAV equipped with a lightweight spectrum analyzer. We demonstrate that the measurements can be highly influenced by the antenna pattern as shaped by the UAV's frame. A calibration procedure is presented to correct for the resulting angular variations in antenna gain. The measurement setup is then validated on real flights from an airstrip at distances in excess of {300}{m}.

\end{abstract}

\IEEEpeerreviewmaketitle

\newcommand{\inches}{\ensuremath{{}^{\prime\prime}}}

\newcommand{\Vasilii}[1]{\noindent 
\textcolor[rgb]{0.2,0.6,0.2} {{$\blacktriangleright$ 
   {\textsf{[Vasilii]: #1}}}}}

\newcommand{\william}[1]{\noindent 
\textcolor[rgb]{0.2,0.2,0.6} {{$\blacktriangleright$ 
   {\textsf{[William]: #1}}}}}

\maketitle

\section{Introduction}
\label{sec:introduction}

There is growing interest in the use of unmamnned aerial vehicles (UAVs) for many connectivity applications where the mobile nature of the UAVs is advantageous as compared with fixed transceivers~\cite{droneReport,ZenGuvZha2020,SaaBenMoz2020}. 
Indeed, the applicability of UAVs for communication purposes has enormous potential, both recreational and in support of the operation of a variety of industries and businesses. In particular, because UAVs can fly over any type of terrain or obstacles, including traffic jams or blocked streets, they can ensure seamless communication to and from users~\cite{Asadpour_MAV_netw}. Also, integration with autonomous driving cars or first-responders can provide additional safety mechanisms in complex environments~\cite{Mezzavilla_PPDR_B5G}. 

The millimeter wave (mmWave) range is an inviting realm for UAV communication because of the enormous bandwidth availability and the possibility of line-of-sight (LOS) transmissions  \cite{zeng2018cellular}. As with any other technology, the design and eventual deployment of UAV communication at mmWave frequencies requires proper channel models, yet the environments and operating conditions of UAVs largely differ from their terrestrial counterparts and hence the corresponding models are not readily applicable. Custom models are necessary and devising these models, in turn, requires large sets of empirical measurements.

Unfortunately, actual measurement of air-to-ground (A2G)
channels in the mmWave band has been tremendously difficult,
primarily due to the strict payload, size, and power limitations
on commercially available UAVs combined with the
massive form factor of the channel sounders typically 
employed in terrestrial applications. The paucity of
A2G measurements at mmWave frequencies
has been one of most significant barriers to deployment 
of aerial systems at these frequencies.

Various studies of aerial channels at mmWave frequencies are available in the literature~\cite{khawaja_survey2019, Cai_UAV_LTE, Schneckenburger_AerialChannel_posit, Zajic_M2M_wireless_channels, Willink_A2G_MIMO_2016, xia_NeuralNetworks_UAV_channel, polese2020experimental}. However, just a few of them rely on low-altitude platforms that can be representative of UAVs at mmWave frequencies.
In~\cite{Khawaja_mmwave_UAV_2018}, the authors provide an analysis of the small-scale temporal and spatial characteristics of A2G radio channels by means of simulation with ray tracing software. In~\cite{Matolac_tvt_part1}, the authors present A2G channel measurements in the L-band ($m${970}~{MHz}) and C-band 
($m${5}{GHz}), conducted from an aircraft flying at altitudes of {580}~{m} and {900}~{m}. Path loss models for the LTE bands are presented in~\cite{Amorim_UAVcomms_over_cellular} and~\cite{Al-Hourani_cellular2UAV}. A  recent study~\cite{Cui_multiFreq_A2G_analysis} reports path loss measurements at several frequencies between 1 and {24}{GHz}. However, there are only a handful of contributions that focus on real-world mmWave aerial measurements, meaning {28}{GHz} and higher. The measurements most similar to ours are those reported for a street-canyon scenario~\cite{vitucci_A2G_2021} at {27}{} and {39}~{GHz}, based on which power azimuth and elevation profiles are derived. 

To overcome the challenges of mmWave A2G
measurements, we present a lightweight setup characterized by:
\begin{itemize}
\item The use of a directional horn
antenna at a ground transmitter, thereby avoiding
the power consumption limitations at the UAV and capitalizing on reciprocity to measure the A2G channel from the ground.
\item The deployment of a lightweight ({282}~{g}) spectrum analyzer
connected to an Intel processor stick ({68}~{g}) 
that enables accurate channel measurements with minimal payload.
\end{itemize}

A relevant insight of our work is the observation that a main source of errors is the uncertainty in the position of the UAV. 
Small variations in angle can cause significant differences in antenna gain and, subsequently, on the path loss, meaning that an error of several meters in the UAV position leads to offsets of several dB in the path loss.

To validate the measurements, experiments were conducted
at {28}{GHz}\footnote{While {28}~{GHz} is formally not part of the mmWave band ({30}~{GHz} to {300}~{GHz}), it is close enough that it is typically subsumed therein.} in an open space---a dedicated airfield---where the ground-truth path loss is known (Friis' law).
Preliminary experiments demonstrate that we can obtain
accurate measurements, opening the door to extensive measurements in more complex environments.
Also, similar equipment can be utilized at even higher frequencies
and the work can further serve to calibrate ray tracing
simulations~\cite{xia_NeuralNetworks_UAV_channel, xia_MultiArray}.

The structure of this article is as follows. Section~\ref{sec:scenario} presents an overview of the measurement setup and a description of the environment. Section~\ref{sec:measresults} introduces preliminary A2G channel measurements and describes the path loss model that they give rise to. Finally, Section~\ref{sec:conclusion} concludes the paper, highlighting possible issues in A2G measurements and discussing plans for subsequent work.

\section{Measurement Equipment}
\label{sec:scenario}

Our setup was developed at VTT, Finland.\footnote{https://www.vttresearch.com/en/ourservices/millimetre-wave-technologies} The measurements were conducted at {28}~{GHz} by a system composed of two separate nodes: (\emph{i}) the transmitter, positioned on the ground, and (\emph{ii}) the receiver, mounted on the UAV~\cite{semkin_A2G_setup}. The tests were conducted at the Nummela airport, in Finland.

\subsection{Transmitter}
\begin{figure}[!t]
\centering
\includegraphics[scale=0.65]{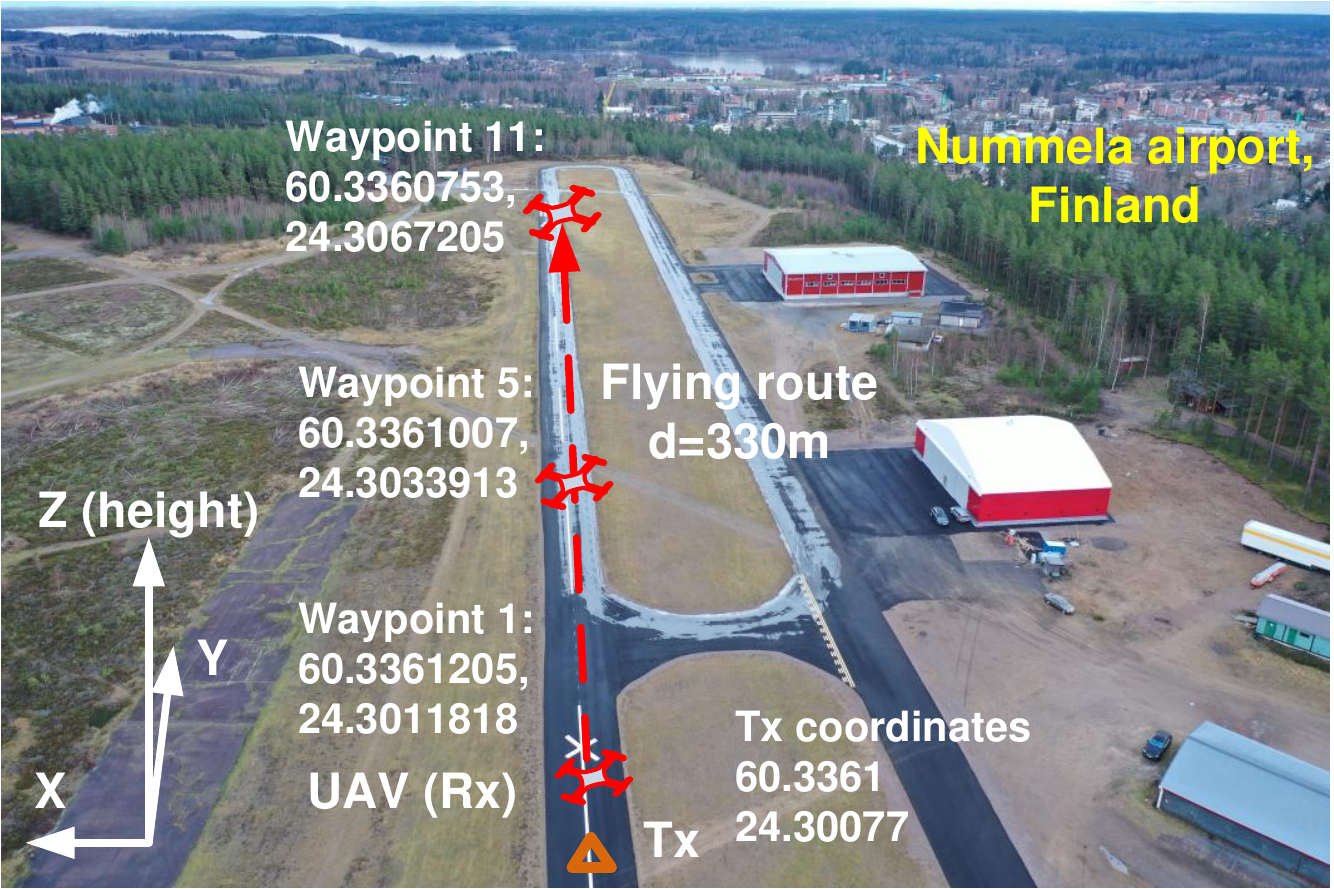}
\caption{Aerial photograph of the measurement location with specified route and coordinates.}
\label{fig:location}
\end{figure}

The signal generator is an Anritsu  MG3697C and the transmitter features an ETS Lindgren 3116 horn antenna with a maximum measured gain of {14.7}~{dBi}. This transmitting antenna was placed on a tripod and an elevation angle of 15$^{\circ}$ was maintained throughout the measurements. The transmit power was set to +{22}~{dBm}. The location and the coordinates of the transmitter are presented in Fig.~\ref{fig:location}.

\subsection{Receiver}
The UAV on which the receiver was installed is a DJI M210 RTK. A compact Intel compute stick PC and an Anritsu spectrum analyzer MS2760A-0100 were mounted on a custom 3D printed structure below the UAV. The spectrum analyzer is also a small-form-factor device with a weights of {282}~{gr}. The equipment was secured with heavy duty velcro straps. The PC was powered by a 2S Li-Po battery through a voltage regulator. Batteries with different capacities, ranging between {2200} and {3000}~{mAh}, were used in the measurement campaign, allowing for the PC to run for 30--40 minutes at an outside temperature of 2$^{\circ}$C. (Batteries with higher capacity could also be featured since the voltage regulator takes up to 40V as an input.) The batteries were chosen to keep the weight of the overall system relatively low. The spectrum analyzer, an external GPS---to record the flight path---and a super-bright LED were connected to the PC via USB-hub; the purpose of the LED was to verify whether the PC was indeed recording during the flight.

Another 4S Li-Po battery acted fed the power amplifier. The receive antenna, omnidirectional in azimuth and having a measured gain of {2.1}{dBi}, was installed on the same 3D printed fixture. In addition, a secondary GPS device was installed on the UAV for more accurate flight recording. The measurement equipment was distributed to balance the UAV and avoid interference between different parts of the system and to ensure that as many GPS satellites as possible were visible. The portable PC was shielded with copper tape.

\begin{figure}[t]
\centering
\subfloat[Receiver block diagram.]
{\includegraphics[scale=0.44]{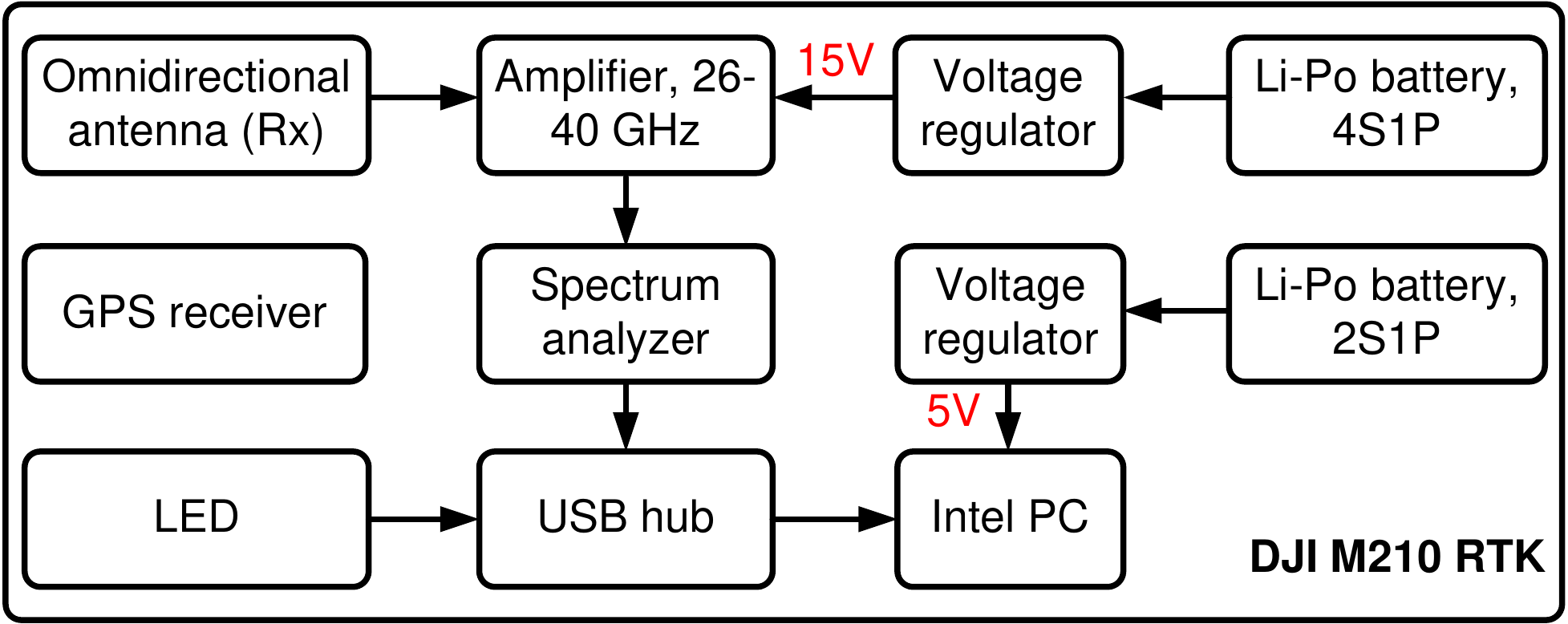}}
\quad
\subfloat[Photograph of the UAV carrying the measurement equipment.]
{\includegraphics[scale=0.57]{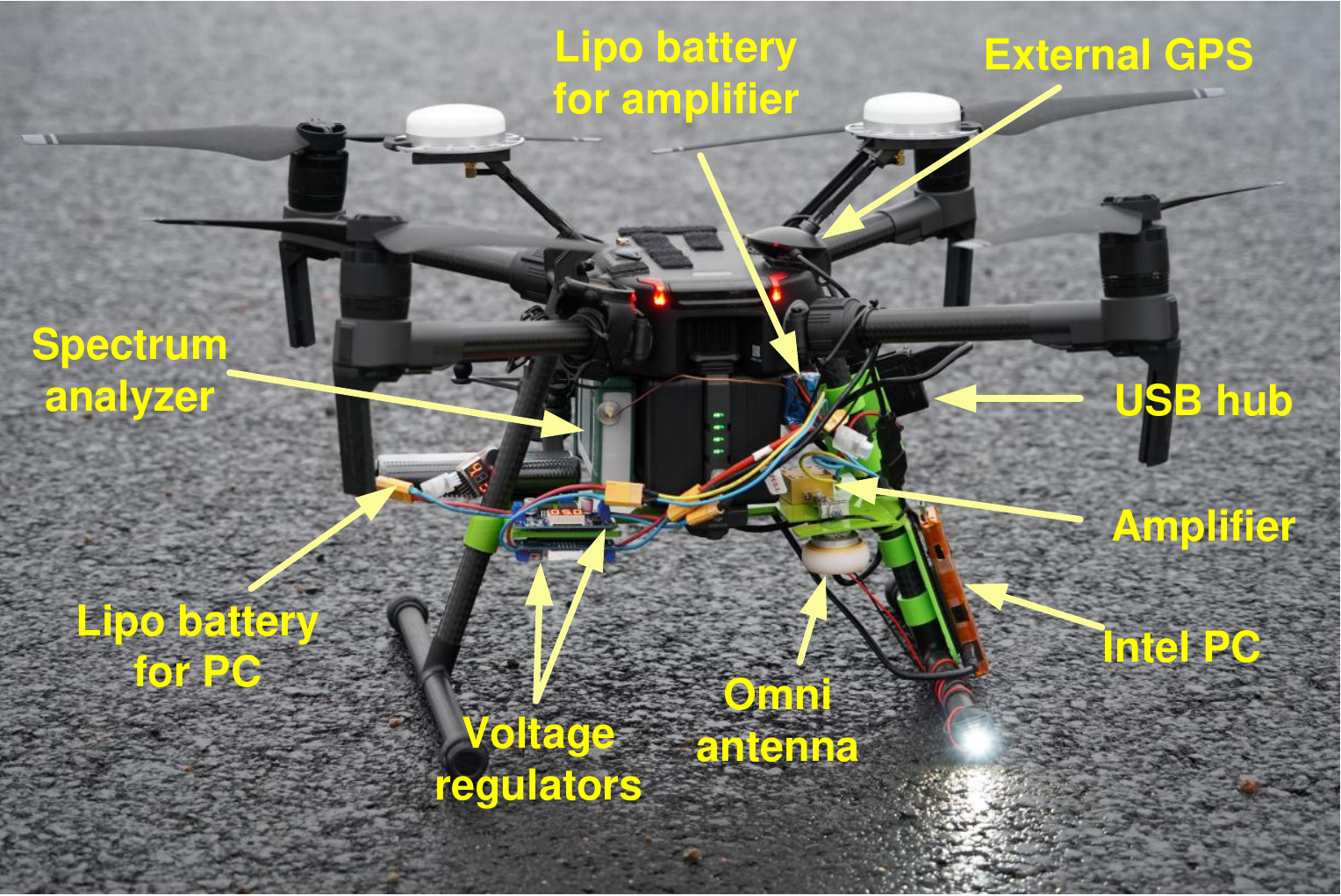}}
\caption{UAV-based measurement setup for power measurements at 28 GHz.}
\label{fig:setup}
\end{figure}

The total weight of the equipment was {1.27}{kg} (see Table~\ref{tab:equiplist}), which is below the maximum payload of {1.5}{kg} that the UAV is able to carry with standard TB50 batteries. The weight of the UAV is {3.8}{kg} while the maximum takeoff weight is {6.14}{kg}. The block diagram of the receiver and a photograph of the UAV can be found in Fig.~\ref{fig:setup}.

\begin{table*}[!t]
\renewcommand{\arraystretch}{1.3}
\caption{Equipment installed on the UAV.}
\label{tab:equiplist}
\centering
\begin{tabular}{|c|l|c|l|l|}
\hline
\textbf{N\textsuperscript{\underline{o}}}   & \textbf{Equipment} & \textbf{Weight (kg)} & \textbf{Model} & \textbf{Notes}    \\ \hline
1  & Spectrum analyzer                      &  0.282 & Anritsu MS2760A-0100   & Frequency range 9 kHz-110 GHz   \\
2  & Omni antenna 28 GHz                     &  0.068 & Eravant SAO-2734030345-28-S1 & 3 dBi gain     \\
3  & Amplifier 26-40 GHz                    &  0.060 & ERZ-LNA-2600-4000-50-2.5  &  Signal gain 50 dB, Power supply 15V   \\
4  & RF cable                               &  0.012 &   & Flexible cable 50 cm (up to 67 GHz) \\
5  & Intel USB stick                        &  0.068 & Core m3 4 GB 64 GB Flash  & Shielded with copper tape    \\
6  & Power cables    &  0.124 &   & Used for PC, USB hub, etc.  \\
7  & Battery for the PC                     &  0.128 & 2S Li-Po 2200mAh   & Any battery from 2S to 6S can be used  \\
8  & Battery for the amplifier              &  0.224 & 4S Li-Po 4000 mAh   & 4S and larger batteries only\\
9  & 3D printed fixtures                    &  0.106   & & Printed with PLA material \\
10  & GPS antenna                           &  0.046 & Globalsat BU-353S4   & Data from this GPS is not used  \\
11 & USB hub                                &  0.038   &  &Used to connect spectrum analyzer to PC + periphery devices\\
12  & Voltage regulator (x2)                &  0.068    & DFR0379 & Adjustable voltage output 4-40V \\ 
13  & External GPS device                            &  0.046 & Polar Vantage M    & Fixed on the UAV's arm  \\ \hline
  & \textbf{Total}                          &  1.270  &  & \\
\hline
\end{tabular}
\end{table*}

\section{Preliminary Results}
\label{sec:measresults}
\subsection{Methodology}

\begin{figure}
\centering
\includegraphics[scale=0.54]{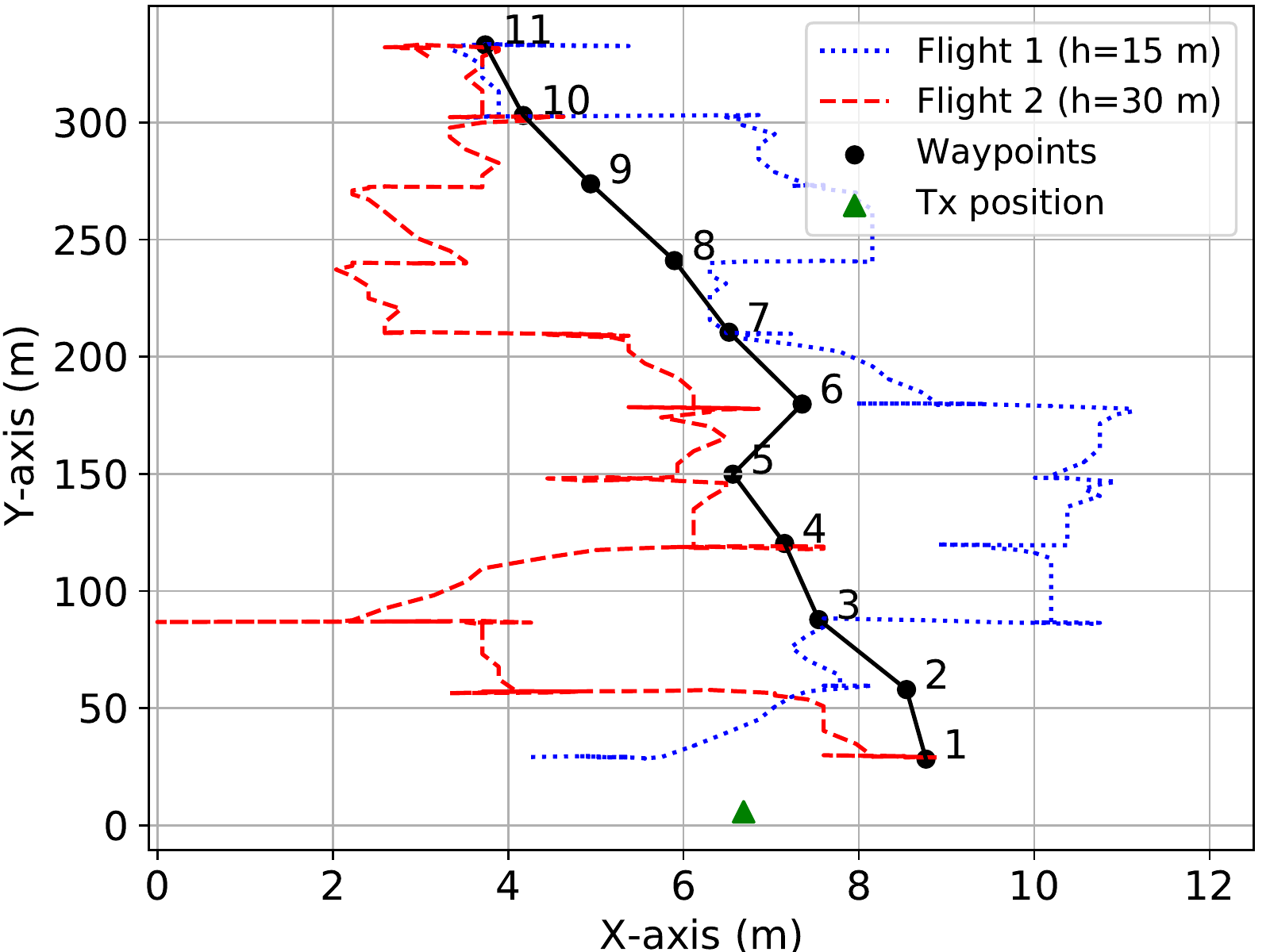}
\caption{Flight route according to the GPS compared with the initially planned route on the XY-plane (height not shown).}
\label{fig:gps}
\end{figure}

In this section, we present the early results gathered at {28}{GHz} and the measured radiation patterns of the antennas that were utilized.
The measurements were conducted at Nummela airport, right above an unused runway as shown in Fig.~\ref{fig:location}. The 11-way point mission indicated in Fig.~\ref{fig:gps} was uploaded prior to the flights and used for each measurement. The UAV would take off in automatic mode, reach the starting point and hover for 30 s, then fly on towards the specified second location and hover thereon for another 30 s, and so on. Once the waypoint mission was complete, the UAV would turn and fly back to the starting point, where it would finally land. Usually, the next flight would be conducted at a different altitude.

The GPS recordings for two flights at respective heights of {15} and {30}~{m} are compared with the corresponding planned waypoint mission in Fig.~\ref{fig:gps}.
Some drifting can be observed in the UAV trajectories with respect to the planned route because of small uncertainties in the GPS receiver and, especially, due to deviations cause by the wind.
The higher altitude flight exhibits a larger deviation from the planned route on account of the more pronounced drifting induced by the wind. When the UAV reaches a desired measurement position, it hovers and tries to maintain that position according to the uploaded GPS coordinates. However, with an increasing height, the wind becomes stronger and these more pronounced shifts from the planned position are observed.

For the data processing, the recorded GPS data was employed and received amplitude values were computed using distances extracted from the GPS data.
The spectrum analyzer recorded 9 samples per second and saved them on the PC, leading to 270 samples per UAV position.
During the post-processing, samples taken while the UAV was in motion towards the starting point, between waypoints, and during ascending/descending maneuvers in-between flights, were all discarded.
The amplitudes of the received signals at each measurement location were averaged to distill the large-scale channel behavior.

\subsection{Antenna Radiation Patterns}
The transmitter and receiver antenna patterns should be carefully determined and taken into account for every calculation. A horn antenna by ETS Lindgren was utilized at the ground transmitter while an Eravant omnidirectional antenna was mounted on the UAV. Antenna measurements were performed in the anechoic chamber of Aalto University, Finland.

\begin{figure}[!t]
\centering
\includegraphics[scale=0.22]{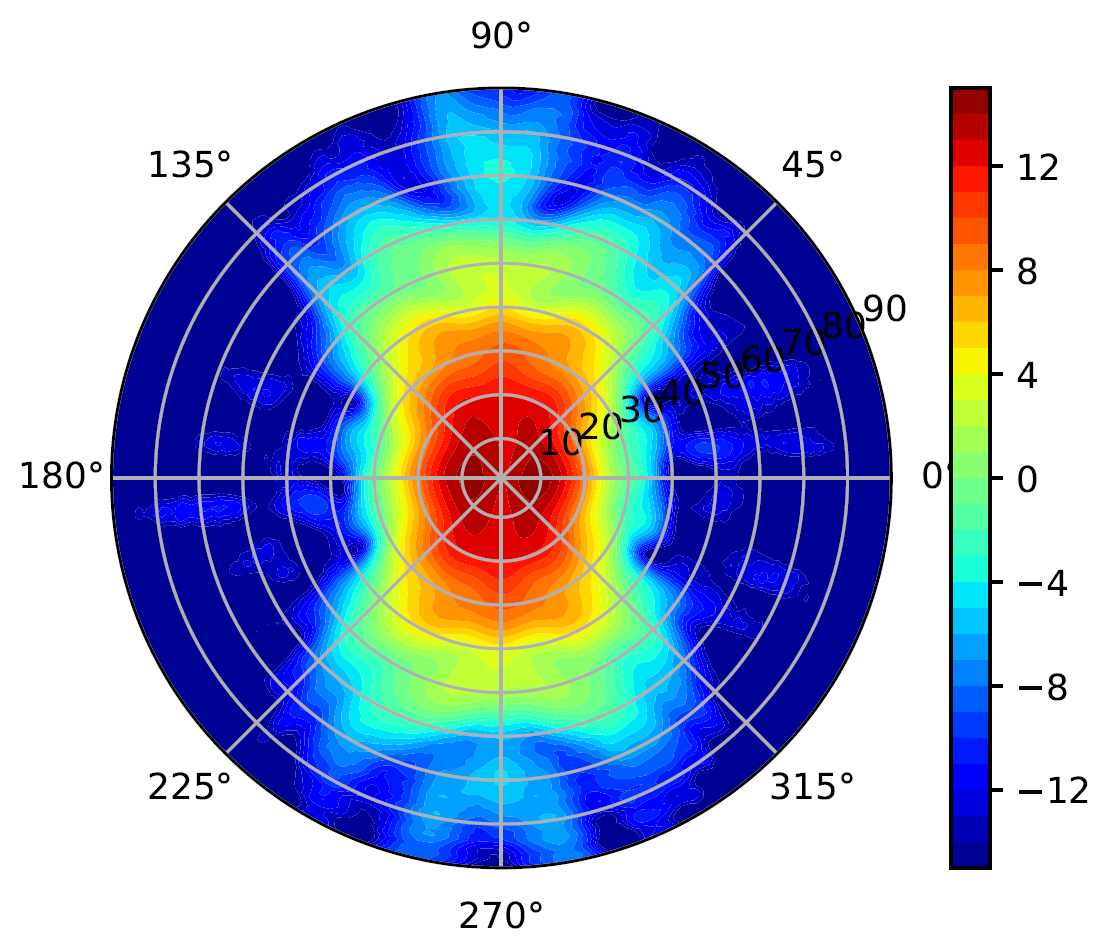}
\caption{Measured radiation pattern of the horn antenna installed on the ground transmitter.
}
\label{fig:horn}
\end{figure}

The measured horn antenna radiation pattern, which exhibits a typical shape and a maximum antenna gain of {14.7}~{dB}, is depicted in Fig.~\ref{fig:horn}.

\begin{figure}[!t]
\centering
\includegraphics[scale=0.42]{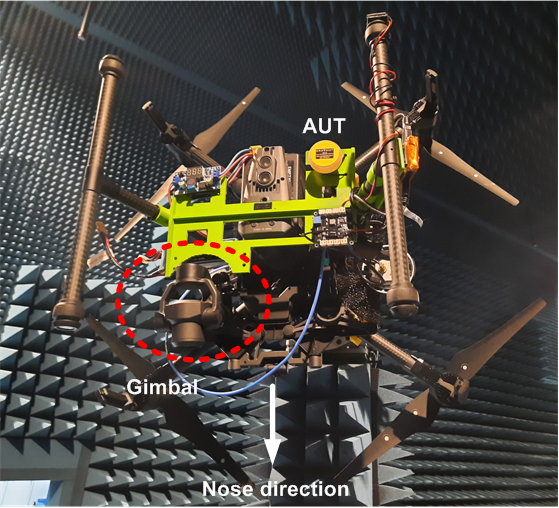}
\caption{Setup for radiation pattern measurements of omnidirectional antenna located on the UAV.}
\label{fig:omni_meas_drone_setup}
\end{figure}

\begin{figure}[!t]
\subfloat[Omnidirectional antenna pattern.]
{\includegraphics[scale=0.22]{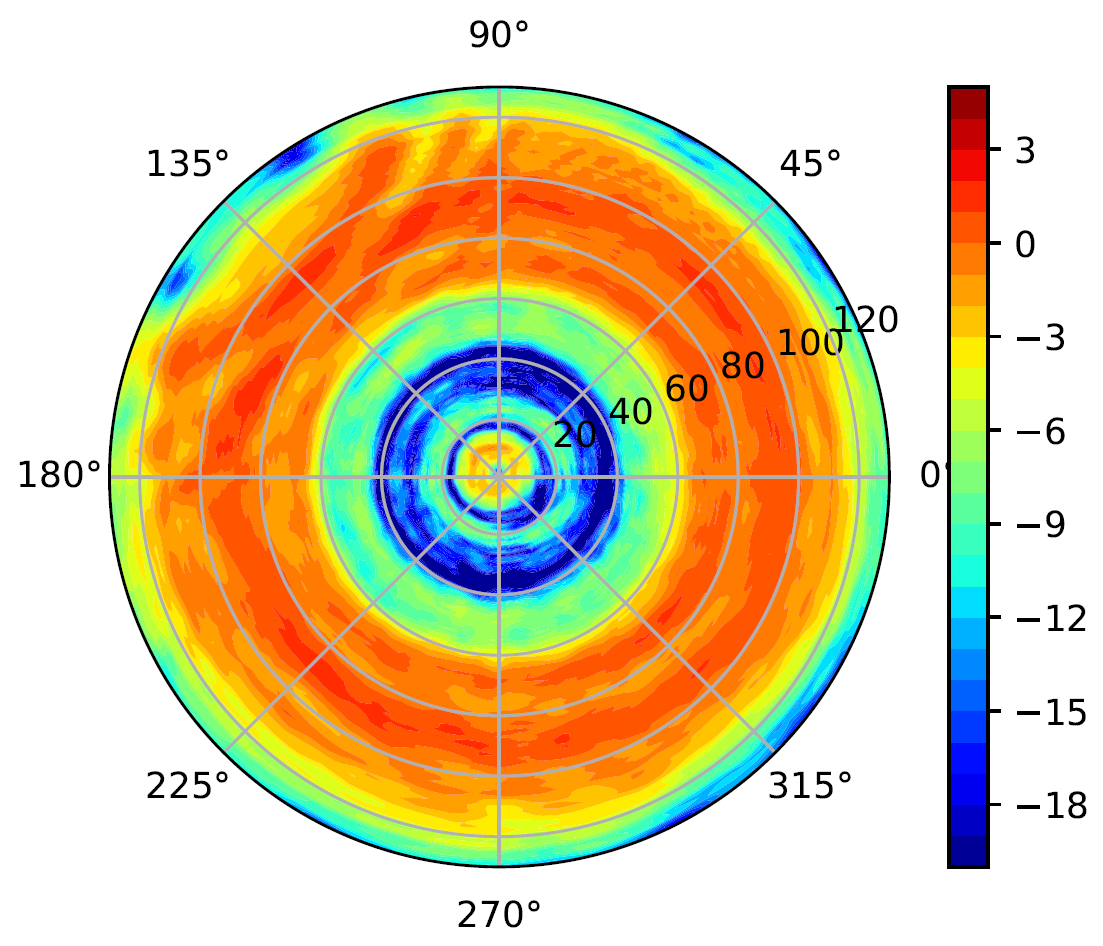}}
\quad
\subfloat[Radiation pattern of the omnidirectional antenna located on the UAV.]
{\includegraphics[scale=0.22]{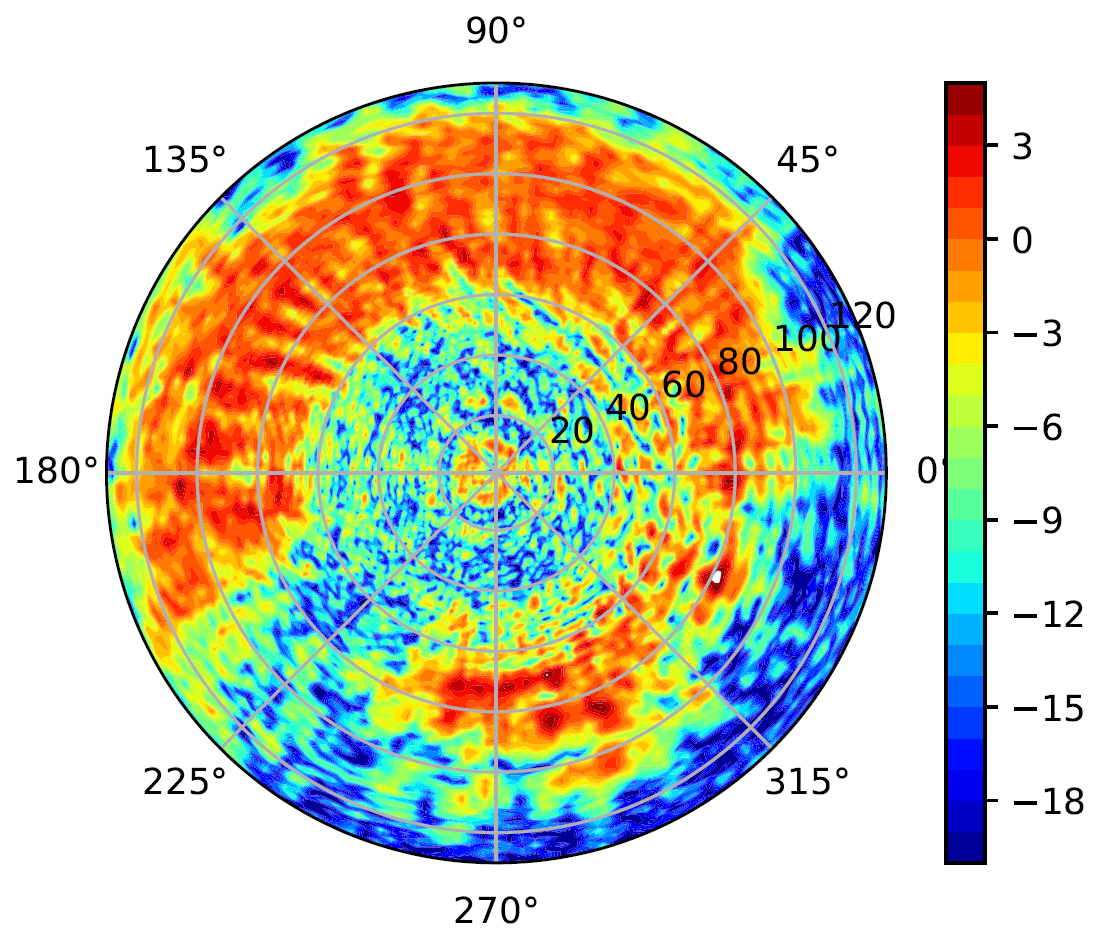}}
\caption{Measured 3D radiation patterns of omnidirectional antenna in free space and installed on DJI M210 as in the measurement setup presented in Fig.~\ref{fig:setup}. Bottom view.}
\label{fig:omni}
\end{figure}

Distinct measurements of the omnidirectional antenna were taken in the anechoic chamber: first, in isolation, and then mounted on the UAV. In the latter case, all equipment was also installed (see Fig.~\ref{fig:omni_meas_drone_setup}) to mimic the actual flight conditions. Fig.~\ref{fig:omni} presents the two measurements. As can be appreciated in Fig.~\ref{fig:omni}(a), the isolated antenna does exhibit a highly uniform radiation pattern on the azimuth plane.
However, as seen in Fig.~\ref{fig:omni}(b), once the antenna is mounted on the UAV, its radiation pattern is severely affected by prominent parts of the UAV's own carbon-fiber hull. For example, the decrease in gain at an azimuth angle of $225^{\circ}$ directly corresponds to the position of the gimbal under the UAV (recall Fig.~\ref{fig:omni_meas_drone_setup}) and indicates that protruding parts of the UAV can affect the antenna radiation patterns. The landing skid of the UAV also has an effect on the antenna pattern. It can be noticed that, at $10^{\circ}$, the pattern is more affected than at $190^{\circ}$, even though the bottom of the UAV is symmetrical. This is due to the close location of the antenna to the right-side landing gear.

The obtained patterns (Figs.~\ref{fig:horn} and~\ref{fig:omni}) are processed by interpolating the data over a mesh grid of azimuth and elevation angles, and used in subsequent calculations. 

\subsection{Path Loss}

The measurement results obtained from two flight at 15 and 30 m were processed and the ensuing path loss model is presented next.

To process all the parameters and extract final path loss values,  we considered the distance vectors between the ground transmitter and the UAV position where the samples were taken. Next, we computed the 3D distances and the azimuth and elevation angles from the distance vectors. Taking into account the transmitter uptilt, we obtained local angles and the ensuing antenna gains. 

Having measured all the losses in the cables, we finally obtained the path loss as
\begin{equation}
 L(d) = L(d_0) + 10 \alpha \log_{10}(d) + \beta,
 \label{eq:logmodel}
\end{equation}
where $d$ is the 3D distance between transmitter and receiver, $d_0=1$ m is a reference distance, $\alpha$ is the path loss exponent, and $\beta$ accounts for all the losses.

The measured power at $d_0$ was {15.4}~{dB}. The measured cable losses between the signal generator and the transmit antenna were {10}~{dB}, the losses in the cable between spectrum analyzer and amplifier were {2.5}~{dB}, and the amplifier gain was {55}~{dB} at {28}~{GHz}. The path loss exponent came out as $\alpha=2.2$ while $\beta=41.6$ dB. The path loss exponent value is a bit higher than that one for free space environment, which is equal to 2, as stated in~\cite{Rappaport_wireless_comms}. Several reasons may cause this difference: i) changes in the yaw while maintaining nose direction and roll angles while hovering which could vary due to strong wind, ii) inaccuracy in the recorded GPS coordinates - this will heavily affect the gain values which should be used in the post-processing, and iii) possibly slightly lower gain of the amplifier due to discharging battery during the measurements, thus leading to smaller amplification. It should be also noted, that we assumed constant height of the UAV during the measurements, however, it was verified that inaccuracy can be in the order of several meters and larger than in the horizontal plane.

\begin{figure}[!t]
\centering
\includegraphics[scale=0.56]{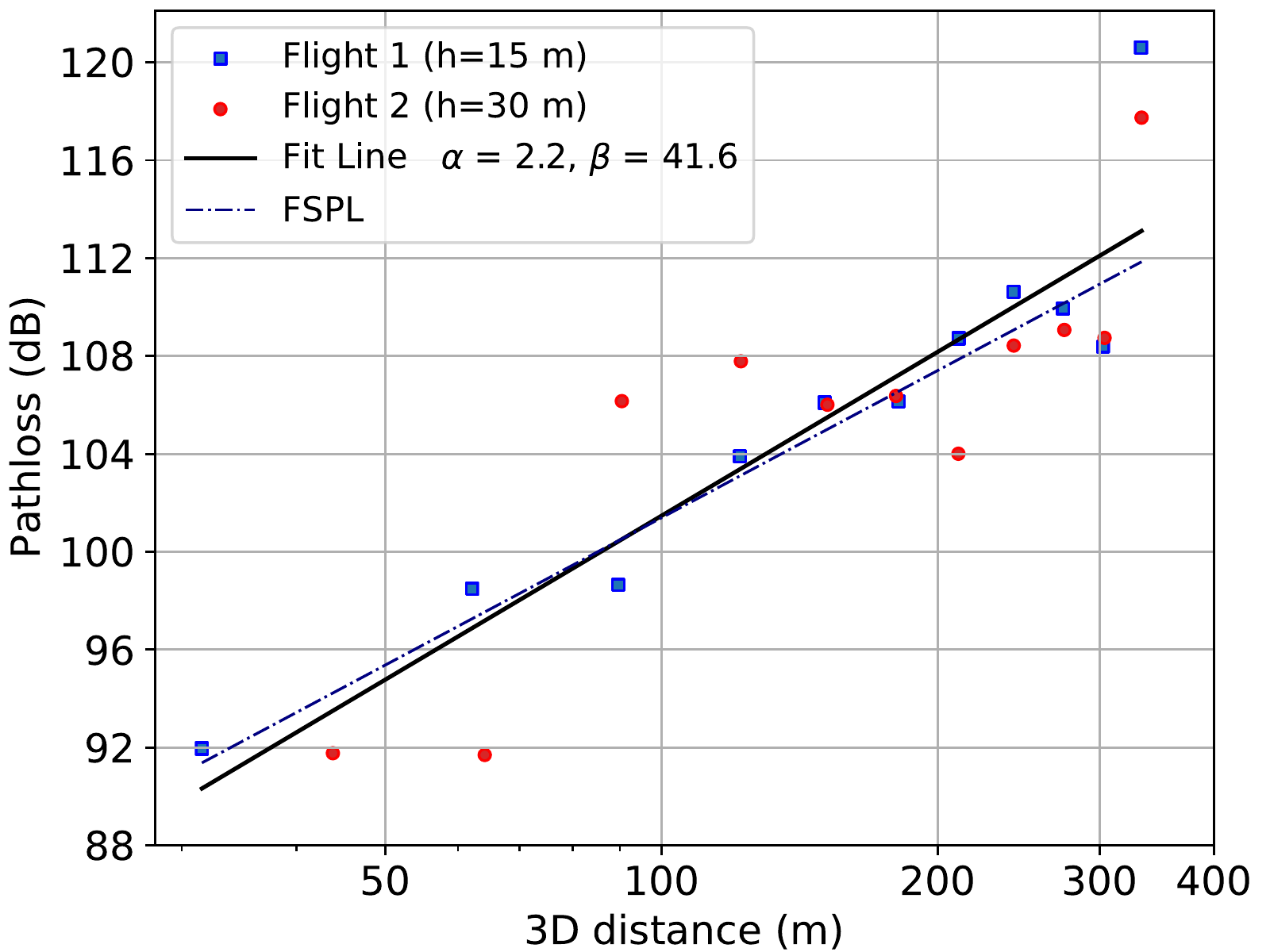}
\caption{Measured path loss values obtained during all flights, and linear fit.}
\label{fig:pathloss}
\end{figure}

\section{Conclusion and discussion}
\label{sec:conclusion}

This work has presented an overview of the UAV-based measurement setup at mmWave frequencies and shown some preliminary results obtained in the open-air field at 28 GHz. Based on the obtained data, a path loss exponent equal to 2.2 was calculated. According to the recorded GPS coordinates of the flights, we verified that there is some drifting of the UAV relative to the planned route. In addition, the antenna patterns were measured and the corresponding antenna gain values were taken into account in the path loss calculations. The measurements performed in this work are relatively simple, but, at the same time, they are extremely time consuming due to many technical and logistical issues. The main purpose of this work was to test the equipment and identify issues that might arise in A2G empirical channel characterizations at mmWave frequencies.

Several sources of errors were pinpointed that should be carefully considered when similar measurement campaigns are executed:
\begin{itemize}
    \item Positioning accuracy of the UAV. Conventional GPS receivers may not be suitable due to potential inaccuracies of several meters. Specifically, any major imprecision in the UAV height should be corrected. One possible solution is utilizing a real-time-kinematic positioning system with the centimeter-accuracy level.
    \item Effect of the UAV's hull on the antenna radiation pattern. As it was shown in this work, the antenna performance can be affected by a particular UAV and antenna location. Therefore, it is important to estimate the effect of the UAV's frame on the antenna patterns.
    \item Verifying yaw and roll angles when the UAV is hovering, and pitch angles when the measurements are conducted in flight. This will help in identifying the correct angles for the antenna patterns.
    \item Wobbling of the UAV due to rotating motors. This may be not so crucial for amplitude measurements, yet it would become critical if full coherent measurements are of interest~\cite{banagar_UAV_wobbling}.
\end{itemize}

For subsequent measurements we are planning to utilize a real-time kinematic system, which will provide higher level of accuracy in terms of positioning of the UAV. We are planning to study wobbling effect on the RF system performance in more details. Our ongoing measurement campaign includes similar measurements at 60 GHz. Next, these measurements will be carried out in urban and sub-urban environments and A2G models will be derived. The results of this research work will enable realizing novel applications and deploying UAV-based communication systems in cities in the nearest future.

\section*{Acknowledgment}

The work of V. Semkin is supported in part by the Academy of Finland.  
The work of W. Xia, S. Rangan, M. Mezzavilla, and G. Loianno is supported by NSF grants  1302336,  1564142,  1547332, and 1824434,  NIST, SRC, and the industrial affiliates of NYU WIRELESS.
The work of A. Lozano and G. Geraci is supported by ERC grant 694974, by MINECO's Project RTI2018-101040, by ICREA, and by the Junior Leader Fellowship Program from ``la Caixa".

\bibliographystyle{IEEEtran}
\bibliography{refs}

\end{document}